\documentclass[letterpaper,times]{sae}

\usepackage{booktabs}
\usepackage[svgnames]{xcolor}
\usepackage{longtable}
\usepackage{pgfplots}
\pgfplotsset{width=10cm,compat=1.9}
\definecolor{SAEblue}{RGB}{1,160,233}
\usepackage{soul}
\usepackage{amsfonts}

\newcommand{\spike}[2]
{\bgroup
  \sbox0{#2}%
  \rlap{\usebox0}%
  \hspace{0.5\wd0}%
  \makebox[0pt][c]{\rule[\dimexpr \ht0+1pt]{0.5pt}{#1}}
  \makebox[0pt][c]{\rule[\dimexpr -\dp0-#1-1pt]{0.5pt}{#1}}
  \hspace{0.5\wd0}%
\egroup}
\usepackage[justification=justified, singlelinecheck=false]{caption}  
\DeclareCaptionFont{eightpt}{\fontsize{8pt}{11pt}\selectfont #1}
\usepackage{lineno}
\usepackage{siunitx}
\usepackage[utf8]{inputenc}
\usepackage{hhline}
\usepackage{float}

\usepackage{array}
\newcolumntype{L}[1]{>{\raggedright\let\newline\\\arraybackslash\hspace{0pt}}p{#1}}
\newcolumntype{C}[1]{>{\centering\let\newline\\\arraybackslash\hspace{0pt}}p{#1}}
\newcolumntype{R}[1]{>{\raggedleft\let\newline\\\arraybackslash\hspace{0pt}}p{#1}}

\usepackage{graphicx}
\usepackage{multirow}
\usepackage{amsmath}

\newcommand{\ignore}[1]{}

\usepackage{nameref}
\usepackage{comment}

\makeatletter
\def\@seccntformat#1{%
  \expandafter\csname c@#1\endcsname\c@section
  }
\makeatother

\usepackage{subfigure}
\usepackage{multimedia}
\usepackage{xcolor}

\PaperTitle{A Latent Space Framework for Modeling Transient Engine Emissions Using Joint Embedding Predictive Architectures}
\usepackage{calc} 

\makeatletter 
\renewcommand\@biblabel[1]{#1. } 
\makeatother

\AddAuthor{Ganesh Sundaram, Tobias Gehra, Jonas Ulmen, Mirjan Heubaum, Daniel Görges, Michael Günthner}{RPTU Kaiserslautern-Landau}

\PaperNumber{2026-01-0423}

\usepackage{float}
\begin{document}
\maketitle
\section{Abstract}

Accurately modeling and controlling vehicle exhaust emissions, particularly during highly transient events such as rapid acceleration, is crucial for meeting stringent environmental regulations and optimizing modern powertrain systems. While conventional data-driven modeling methods, such as Multilayer Perceptrons (MLPs) and Long Short-Term Memory (LSTM) networks, have improved upon earlier phenomenological or physics-based models, they often struggle to capture the complex nonlinear dynamics of emission formation. These monolithic architectures attempt to learn from all available data, which increases their sensitivity to dataset variability. They often require increasingly deep and complex architectures to improve performance, thereby limiting their practical utility. This paper introduces a novel approach that overcomes these limitations by modeling emission dynamics in a structured latent space. Using a rich dataset combining real-world driving data from a Portable Emission Measurement System (PEMS) with high-frequency hardware-in-the-loop test bench measurements, a Joint Embedding Predictive Architecture (JEPA) is leveraged. This framework learns to abstract away irrelevant information and encode only the key factors governing emission behavior into a compact, robust latent representation. The resulting model demonstrates superior data efficiency and predictive accuracy across diverse transient regimes, exhibiting stronger generalization than the high-performing LSTM baseline. Structured pruning and post-training quantization are applied to the JEPA framework to enhance the model's suitability for real-world deployment. This combined strategy significantly reduces the model's computational footprint, minimizing inference time and memory demand, with only a marginal impact on accuracy. This yields a highly accurate model well suited to on-board implementation of advanced control strategies, such as model predictive control or model-based reinforcement learning, in both conventional and hybrid electric powertrains. The results indicate a clear pathway toward more efficient and robust emission control systems for next-generation vehicles.

\section{Introduction}

\subsection{Motivation}

The critical requirement to reduce combustion engine emissions has driven substantial innovation in powertrain technology, particularly with the rise of hybrid electric vehicles. These systems deliver significant fuel-efficiency gains, enabling low-emission operation across various operating modes. However, optimizing interactions between the combustion engine and the electric motor remains challenging, especially under transient conditions, to minimize fuel consumption and regulated emissions such as NO\(_x\), HC, and CO. The requirement becomes increasingly stringent as regulations mandate low emissions across standardized cycles and real-world transient operations. Hence, accurate real-time prediction of engine-out emissions is fundamental to addressing this challenge. 

The predictive models serve two critical roles. First, during the development and calibration of new powertrains, a digital twin is created to accelerate design iterations, reduce reliance on costly physical testing, and enable virtual screening of control strategies. This enables engineers to optimize system performance and potentially reduce the cost and complexity of aftertreatment systems. Second, these models allow advanced predictive control strategies when directly embedded in a vehicle's electronic control unit (ECU). This enables the powertrain to proactively adjust parameters such as air-fuel ratio, spark timing, and torque split, thereby mitigating emissions peaks before they occur and ensuring robust compliance in real-world driving without compromising efficiency.

Therefore, the primary motivation of this work is to develop an improved emission model, compared with earlier models, that is not only highly accurate under transient conditions but also computationally efficient for on-board implementation. The goal is to develop a model that can operate solely on the signals available on production ECUs, bridging the gap between high-fidelity simulation and practical real-time control, and thereby enabling cleaner, more efficient next-generation vehicles.

\subsection{Problem Statement}

While conventional physics-based emission models can adequately represent vehicle behavior under steady-state or quasi-static conditions, their predictive fidelity declines significantly during transient operations. The formation of emission species is governed by complex, time-dependent thermochemical phenomena characterized by strong nonlinearities and coupled dynamics that are difficult to parameterize explicitly. This fundamental limitation has motivated the shift toward data-driven methodologies, particularly Deep Neural Networks (DNNs), which can learn these complex input-output relationships directly from empirical data, thereby implicitly capturing various dynamics in the emission system without requiring explicit model formulation.

Despite their advantages, the widespread use of monolithic neural network architectures, such as Multi-Layer Perceptrons (MLPs) and Long Short-Term Memory (LSTM) networks, introduces challenges. While recursive architectures, such as LSTMs, have demonstrated superior performance, a conceptual deficiency inherent in this approach has been identified. These monolithic models aim to establish a direct mapping from the entire input space to the output space, forcing the network to learn a highly complex function that struggles to separate causal signals from background noise and spurious correlations. Consequently, improving model accuracy often requires increasing network depth and parameter counts, thereby increasing computational demands and the risk of overfitting.

This architectural problem presents two primary issues. First, the models exhibit limited generalization because their performance is tied to the statistical distribution of the training data. Second, when constrained to a low-dimensional input space, the model's output is typically disproportionately sensitive to small perturbations in the input. The primary issue, therefore, is that monolithic models approximate system dynamics rather than distill a robust underlying representation of the system's governing principles. This necessitates a paradigm shift toward an architecture that learns a disentangled, low-dimensional latent representation of the core emission dynamics. This provides a more generalizable and computationally tractable solution for real-world predictive control.

\subsection{Contributions}

To address the limitations of traditional monolithic neural-network approaches, this work investigates Joint Embedding Predictive Architectures (JEPA) for modeling transient dynamics of gaseous emissions. In a JEPA, the model does not predict future emissions directly from raw input histories. Instead, it learns a joint embedding of past and (masked) future variables in a common latent representation space. It is trained to bring embeddings corresponding to physically compatible past–future pairs closer together, whereas incompatible pairs are pushed apart. In this work, we use the term latent-space model to denote architectures that operate on such learned internal representations: the raw input and state variables are first encoded into a latent space, and the temporal evolution of emissions is modeled in that space rather than directly in the original measurement space. This encourages the network to focus on a physically meaningful and control-relevant structure in the system dynamics, rather than approximating the full nonlinear input–output mapping in a high-dimensional signal space, as in standard sequence-based models (including NARX-type and recurrent architectures).

For a consistent and rigorous evaluation, the same dataset and LSTM network from prior work~\cite{sundaram2023modeling}, where it outperformed feed-forward and other shallow learning algorithms, are reused. The JEPA model is developed based on the conceptual principles outlined in~\cite{ulmen2025learning} and is trained on emission data collected under real driving conditions and controlled test-bench experiments. A comparative analysis between the JEPA and LSTM models is performed to quantify improvements in transient emission prediction, particularly in capturing rapid variations, peak amplitudes, and temporal phase shifts.

Finally, post-training compression techniques such as structured pruning and post-training quantization are applied to the trained JEPA model to enable efficient deployment on embedded control hardware. The resulting trade-offs between accuracy, inference speed, and memory footprint are systematically analyzed. The combined latent-space representation and compression framework proposed in this study provides an interpretable, computationally efficient, and high-fidelity modeling approach, suitable for accurate emission prediction and advanced control applications in both hybrid and conventional powertrains.

\section{Related Work}

\subsection{Data-Driven Emission Modeling}

Data-driven approaches to engine emission modeling have expanded rapidly, offering a compelling complement to physics-based methods, particularly for transient operations in which nonlinearities, transport delays, and hysteresis predominate. Ensemble techniques, such as Random Forests, have demonstrated strong performance for particulate prediction in boosted GDI applications. Broader comparisons have likewise highlighted Gradient Boosting and XGBoost as among the most accurate supervised learners for emission forecasting~\cite{akehurst2021random, Ding2024}. In parallel, deep architectures ranging from DNNs and CNNs to recurrent models have been widely adopted for gaseous species, reflecting their capacity to approximate complex, multivariate emission dynamics. The primary focus of recent work is the comparative evaluation of feedforward and sequence-aware networks for transient tasks. Multiple studies report that feed-forward neural networks and recurrent models can both achieve high accuracy for regulated species such as NO\textsubscript{x} under drive cycles, while consistently finding that sequence-aware models, especially LSTMs, outperform feed-forward baselines on transients due to their ability to learn temporal dependencies relevant to aftertreatment dynamics and in-cylinder state evolution~\cite{sundaram2023modeling, shin2021predicting}. Extensions that couple LSTM backbones with meta-learning further enhance data efficiency and robustness in low-data or distribution-shifted regimes, reinforcing the advantage of temporal representation learning for rapidly changing operating conditions~\cite{Li2024}. Input selection remains a critical determinant of fidelity and generalization. While CO\textsubscript{2} is often predictable from compact feature sets (e.g., speed–load proxies), accurate modeling of NO\textsubscript{x} and unburned hydrocarbons typically requires richer descriptors capturing air-path, combustion, and thermal states, as well as features sensitive to exhaust transport and catalyst storage/release processes~\cite{huang2022use,fang2021application}. This induces a practical trade-off between predictive performance and instrumentation burden, motivating variable-importance analysis and feature-economical surrogates to manage sensor cost and calibration complexity~\cite{huang2022use,fang2021application}.

Hybrid schemes integrate learning modules with a phenomenological structure to balance accuracy, interpretability, and extrapolation under off-cycle dynamics. Such approaches leverage physics where it is strong and machine learning where it is reasonable, yielding improved transient prediction and better control-oriented behavior~\cite{gehra2025ki}. For instance, fusing CFD-generated fields with learned surrogates (e.g., CNN-based mappers) improves predictions for challenging emission species, such as CO, HC, and smoke, while retaining mechanistic cues from high-fidelity solvers~\cite{Warey2023}. These advances, while effective, still primarily rely on direct input–output mappings, which motivates the present focus on representation learning that targets temporally causal, physically meaningful latent states to improve robustness, data efficiency, and downstream controllability.

\subsection{Latent State-Space Modeling}

Latent state-space approaches learn compact coordinates in which nonlinear dynamics are easier to predict and estimate, enabling multi-step forecasting, observer design, and control with reduced computational costs. A key thread uses encoder–decoder models to identify regions where dynamics are approximately linear, enabling the application of classical linear tools for prediction and control~\cite{Lusch2018Koopman}. Complementary sparse system identification directly discovers low-dimensional coordinates and governing equations from data, improving interpretability and facilitating structure-preserving control synthesis~\cite{Champion2019SINDy}. For controlled systems, Koopman formulations with explicit actuation clarify how inputs enter latent evolution, which is critical when pairing learned models with feed-forward strategies and predictive control~\cite{Proctor2018KIC}.

In automotive and powertrain settings, latent lifting has been explored to create fast, control-oriented surrogates of strongly nonlinear subsystems across various operating regimes. Such predictors can improve transient fidelity relative to black-box baselines while maintaining efficient inference within embedded deployment budgets. More broadly, reward- and task-centric latent models from model-based reinforcement learning reduce representational burden by encoding only control-relevant factors, enabling planning over learned dynamics in compact latent spaces~\cite{Hansen2022TD-MPC, Ji2023DualPolicy, Lin2024TD-MPC2}. These ideas focus on controllers with latent state representations: the learned coordinates serve as the computational bottleneck for downstream control, so representations must balance compactness, temporal consistency, and identifiability to remain compressible without eroding closed-loop behavior.

\subsection{Model Compression Strategies}

Compression for control-focused neural networks encompasses pruning, quantization, low-rank factorization, and knowledge distillation. However, deployment constraints render the choice of technique context-dependent. Surveys and empirical studies consistently show that structured pruning yields practical speedups by reducing tensor sizes while preserving dense computation. In contrast, unstructured sparsity often needs specialized kernels to realize latency gains~\cite{Cheng2017Survey, Han2016DeepCompression}. Dependency-graph-based structured pruning ensures dimensionally consistent group removal across layers and components, which is especially important for multi-module control architectures with encoders, dynamics, and policy/value heads~\cite{Fang2023DepGraph}. Low-rank and tensor factorization compress large linear and convolutional operators while maintaining accuracy after brief fine-tuning, offering hardware-friendly acceleration on general-purpose CPUs/GPUs~\cite{Denton2014Linear, Kim2016Compression}.

Quantization can deliver substantial memory and bandwidth savings, and with adequate hardware support or quantization-aware training, it can preserve accuracy and latency targets~\cite{Jacob2018Quantization, Krishnamoorthi2018Quantizing}. Knowledge distillation provides strong baselines for a student model, a network trained to imitate a larger teacher, while keeping the architecture fixed. It remains a reliable complement to pruning or factorization in control pipelines~\cite{Hinton2015Distill}. For controllers with latent models, two main findings exist: first, careful post-compression fine-tuning is crucial for recovering performance and preserving control-theoretic margins; second, hybrid pipelines that moderate structured pruning or factorization, combined with quantization, often guided by component-aware dependency graphs, achieve the best trade-offs under tight runtime budgets~\cite{sundaram25_appspecific}. Taken together, effective compression for latent-space controllers should preferably have a hardware-aligned structure (structured sparsity or low-rank), retain encoder fidelity to avoid latent distortion, and incorporate fine-tuning and task-level checks to safeguard closed-loop stability and performance~\cite{Han2016DeepCompression}.

\section{Experimental Setup}

\subsection{Emission Dataset}
\label{subsec:emissiondataset}

The applicability and quality of any data-driven emission modeling approach depend critically on the comprehensiveness and fidelity of the underlying dataset. A specialized dataset was generated using a two-stage measurement process established in previous work to ensure comprehensive coverage of real-world operating conditions, with a focus on robust transient emissions modeling. A detailed discussion of the dataset design, measurement configurations, and data preprocessing procedures is provided in~\cite{sundaram2023modeling}. The dataset was created in two steps. First, on-road measurements were conducted with a BMW 530e equipped with a Portable Emissions Measurement System (PEMS) to capture emission-relevant load profiles. The focus was on real driving situations in urban areas, on rural roads, and on motorways. The test vehicle was a HEV that employed different operating modes, ensuring that the combustion engine was operated across a wide range of its engine map (e.g., in battery-sustaining mode) and under transient conditions. Second, these recorded profiles were reproduced under controlled conditions on an engine test bench using the same BMW B48 engine. Additionally, the entire engine map was measured on the test bench using 66 evenly distributed steady-state operating points.

\begin{figure}[h]
  \centering
  \includegraphics[width=\columnwidth]{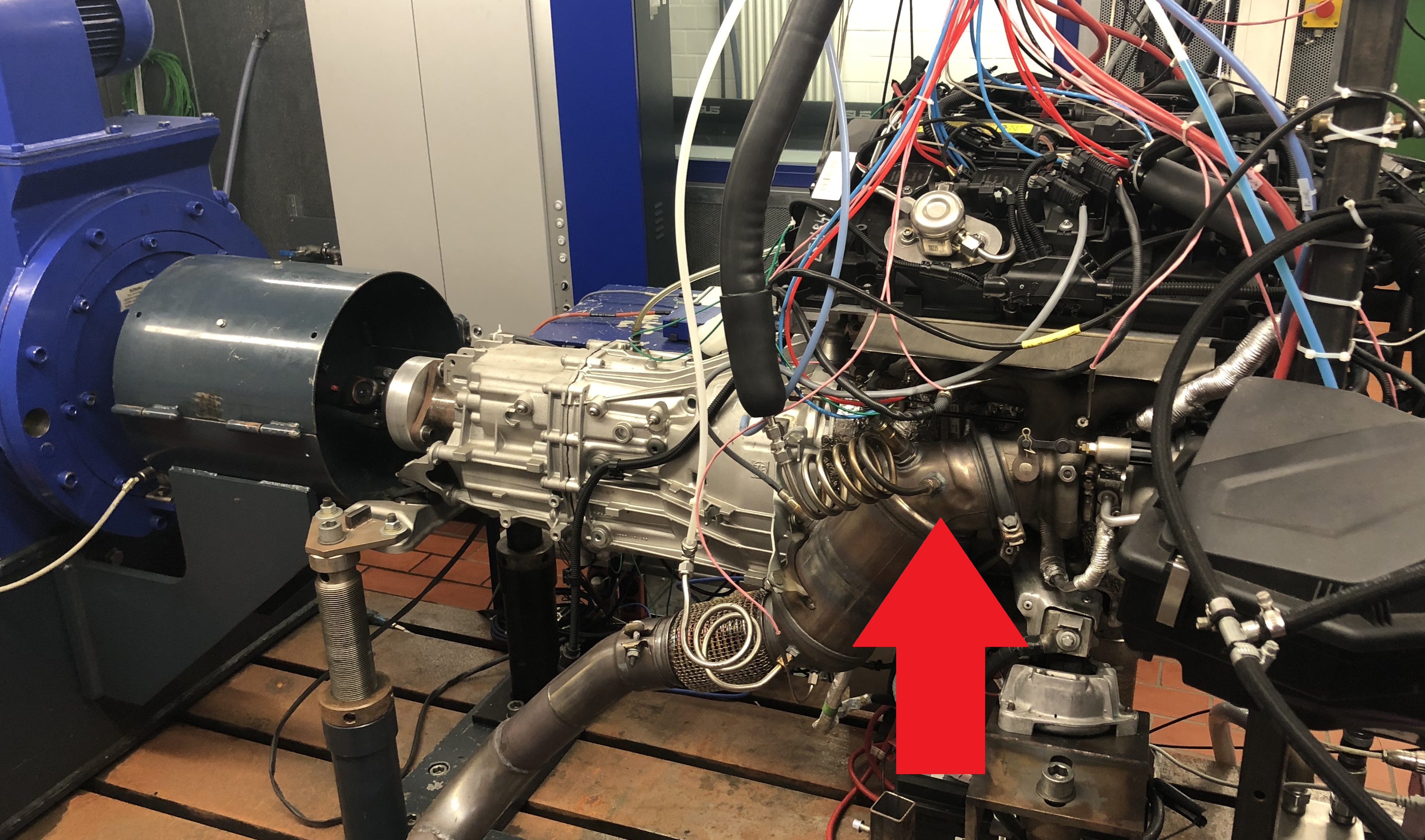}
  \caption{Test bench for dataset generation. \\
  Raw exhaust is sampled between the turbocharger and the catalytic converter (red arrow), allowing for transient-resolved emissions measurements upstream of aftertreatment systems.}
  \label{fig:test_bench_real}
\end{figure}

\begin{figure}[h]
  \centering
  \includegraphics[width=\columnwidth]{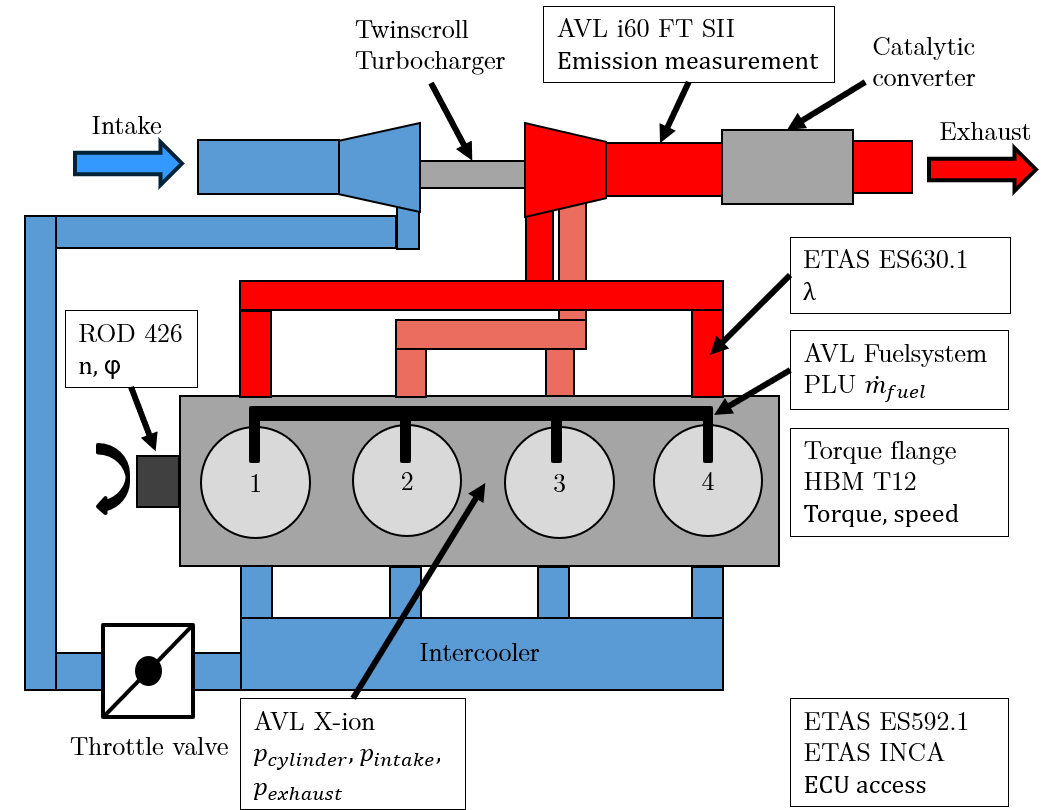}
  \caption{Schematic of the test bench setup. \\
  The engine is coupled to an electric machine (load) via the torque flange and a central measurement system. Standard thermocouple and pressure sensors are omitted here for clarity.}
  \label{fig:test_bench_setup}
\end{figure}

This procedure enables reproducibility and access to richer, bench-level measurements. An illustrative on-road speed–torque profile is shown in Figure~\ref{fig:speed_torque_profile}. The corresponding engine and bench setup used for emission and process data acquisition are depicted in Figure~\ref{fig:test_bench_real} and~\ref{fig:test_bench_setup}, respectively. Table~\ref{tab:b48_specs} summarizes the engine technical specifications. 

Exhaust-gas analysis was performed using Fourier-transform infrared (FTIR) spectroscopy at a sampling rate of \unit{5}{Hz}. In total, \num{146} measurement channels were recorded over more than \num{10} hours of bench testing, establishing a broad empirical basis for subsequent modeling efforts. This dataset serves as the empirical foundation for the modeling, training, and evaluation of all models presented in this study.

\begin{figure}[h]
  \centering
  \includegraphics[width=\columnwidth]{images/figure_1.jpg}
  \caption{On-road speed–torque profile for dataset generation. \\ The envelope and transient excursions derived from this profile are used to define bench trajectories and to focus measurements in emission-critical regions.}
  \label{fig:speed_torque_profile}
\end{figure}

\renewcommand{\arraystretch}{1.2}
\begin{table}[h]
  \centering
  \caption{Key specifications of the BMW B48 four-cylinder gasoline engine used on the test bench as the reference system for transient emission modeling experiments.}
  \label{tab:b48_specs}
  \begin{tabular}{p{0.5\linewidth}p{0.15\linewidth}p{0.15\linewidth}}
    \toprule
    \textbf{Engine parameter} & \textbf{Value} & \textbf{Unit} \\
    \midrule
    Maximum power     & 135   & \unit{kW} \\
    Maximum torque    & 290   & \unit{Nm} \\
    Displacement      & 1.998 & \unit{dm^{3}} \\
    Compression ratio & 11.0  & -- \\
    Bore              & 82.0  & \unit{mm} \\
    Stroke            & 94.6  & \unit{mm} \\
    \bottomrule
  \end{tabular}
\end{table}

\vspace{-0.2cm}

\subsection{Data-Driven Emission Modeling Framework}
\label{subsec:datadrivenemissionmodeling}

The primary objective of this study is to predict the emission values of NO\(_x\), CO\(_2\), CO, and THC, collectively denoted as the output \(X(t)\), from an internal combustion engine (ICE). The model uses a set of input parameters, \(U(t)\), which include engine torque \(\tau\), engine speed \(\omega\), and the air-fuel ratio \(\lambda\). These parameters are derived from the emission dataset described in the earlier section and are used because these signals are readily measurable and integrate directly into hybrid operating strategy optimization. Previous work~\cite{sundaram2023modeling} proves the superior performance of LSTM networks over traditional Multi-Layer Perceptrons (MLPs) for this specific prediction task. 

Consequently, this work investigates whether a state-of-the-art JEPA can surpass the predictive accuracy of the established LSTM baseline. To facilitate this comparative analysis, both an LSTM and a JEPA model were developed and trained on a workstation equipped with an AMD Ryzen Threadripper \num{3960}X CPU (\num{24} cores, \num{3.8}{GHz} base frequency), \num{64}{GB} RAM, and an NVIDIA RTX A\num{5000} GPU. All inference was conducted in PyTorch's eager mode, as elaborated in the subsequent subsections.

Models are compared using the (weighted) mean squared error (MSE), where each emission species is evaluated individually on the same scale after min--max normalization to $[0,1]$; using the weighted MSE, we can additionally emphasize specific emission features via $w_i$, whereas the underlying paper scores all species equally by setting $w_i=1$ for all $i$.
\begin{equation}
\mathrm{WMSE}
= \frac{\sum_{k=1}^{K} w_k \left(\frac{1}{T}\sum_{t=1}^{T}\bigl(X_k(t)-\widetilde{X}_k(t)\bigr)^2\right)}{\sum_{k=1}^{K} w_k},
\qquad w_k \ge 0
\end{equation}

\subsubsection{Baseline LSTM Model Architecture}

The LSTM model from the prior study~\cite{sundaram2023modeling} is characterized as a monolithic architecture. This designation implies a single end-to-end network that directly maps the input sequence \(U(t)\) to the predicted emission concentrations \(\widetilde{X}(t)\). The model processes an input tensor with dimensions [timesteps, features], thereby simplifying the training pipeline. The training process aims to minimize a loss function that quantifies the discrepancy between the predicted emissions \(\widetilde{X}(t)\) and the ground-truth values \(X(t)\). The key hyperparameters from the optimization study, along with the performance characteristics of the baseline LSTM model, are summarized in Table~\ref{tab:lstm_training}. For later model architecture comparisons, it is worth noting that, given the data volume, the LSTM performs better with a comparatively small network, and enlarging the network does not significantly improve performance. 

\begin{table}[h]
\centering
\begin{tabular}{p{0.6\linewidth}cp{0.3\linewidth}} 
\toprule
\textbf{Parameter} & \textbf{Value} \\
\midrule
    Neurons per layer (width)           & 512 \\
    Number of layers (depth)            & 3 \\
    Number of timesteps                 & 10 \\
    Loss                                & Weighted MSE \\
    Model size                          & \SI{2.02}{\mega\byte} \\
    Inference time (total)              & \SI{0.4063}{\second} \\ 
    Inference time (per timestep)       & \SI{0.2032}{\milli\second} \\
\bottomrule
\end{tabular}

\caption{Configuration and empirical characteristics of the LSTM model during training. 
The network architecture is defined by the number of layers (depth) and the number of neurons per layer (width). The model processes sequences of ten timesteps and is trained using a weighted mean squared error (MSE) loss function. The \textit{total inference time} refers to the total time required to perform inference of \num{2000} timesteps, and the \textit{per-timestep inference time} indicates the average time taken for each timestep during inference.}
\label{tab:lstm_training}
\end{table}

\subsubsection{Joint Embedding Predictive Architecture (JEPA)}
\label{subsubsec:jepa}

\begin{figure*}[t]
    \centering\includegraphics[width=0.95\textwidth]{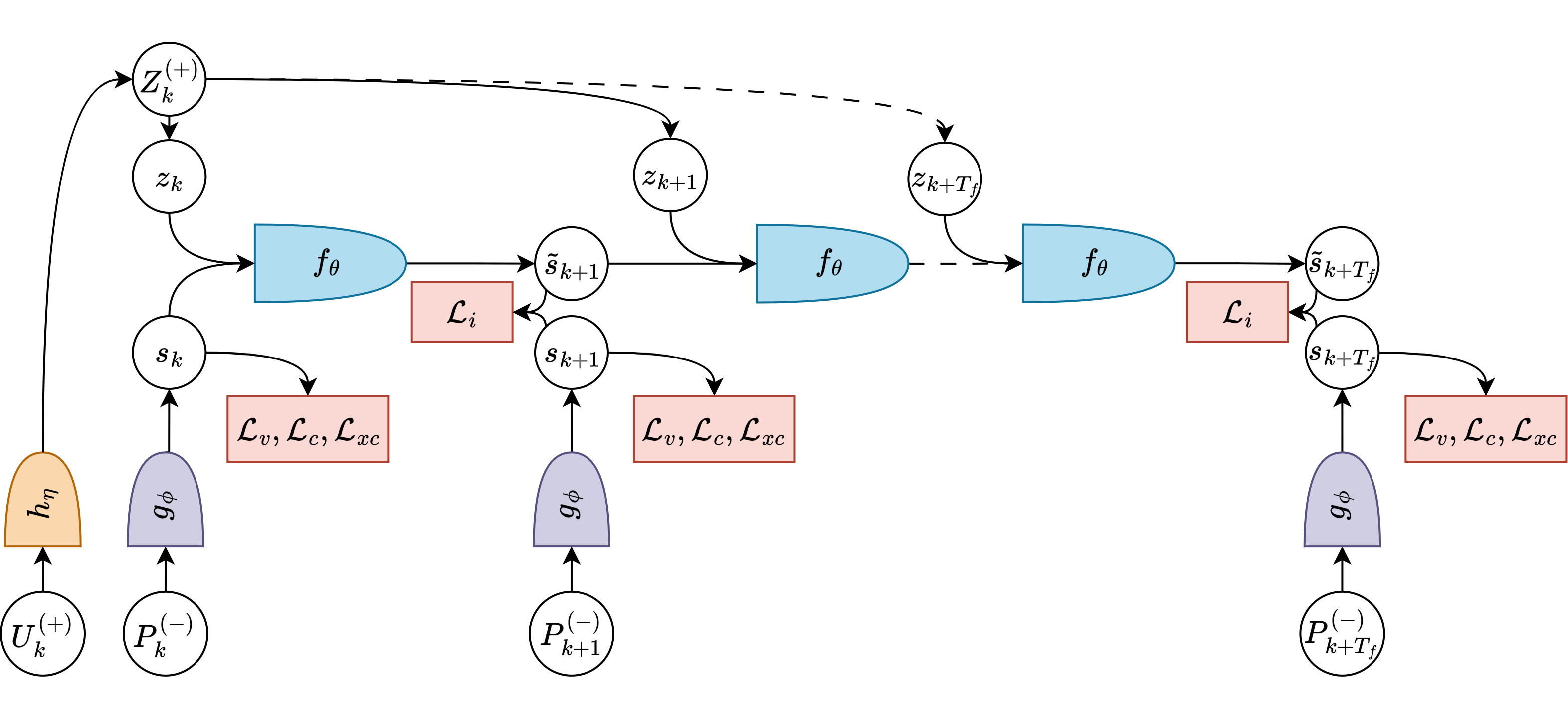}
     \caption{
        Schematic representation of the JEPA training and inference process.\\
        Illustrates how future input sequences and past emission observations are jointly encoded into a latent space. The model predicts future latent states via autoregressive dynamics, using multiple loss functions (variance, invariance, covariance, and cross-covariance) to regularize latent representations. Encoders, predictor network, and losses are highlighted for clarity. Arrows denote the sequence of information flow through encoders $h_\eta$, $g_\phi$, and the predictor $f_\theta$ for emission forecasting across time horizons.
    }
    \label{fig:jepastructure}
\end{figure*}

The proposed JEPA framework comprises two dedicated encoders and a predictive latent-dynamics model, as depicted schematically in Figure~\ref{fig:jepastructure}. JEPA maps both emission inputs and measurements from their original domains into a structured latent space, enabling more efficient modeling of the temporal evolution of emission concentrations. For supervised learning, the dataset consists of emission vectors \( p_k \in \mathbb{R}^{m_p} \) and input vectors \( u_k \in \mathbb{R}^{m_u} \) at discrete sampling indices \( k \in \mathbb{N}_0 \), where \( t_k \) denotes the sampling instant.

\textbf{Latent Space Encoders:} The future input sequences \( U_k^{(+)} \) are projected into the latent domain by an input encoder
\begin{equation}
    h_\eta \big(U_k^{(+)} \big) = Z_k^{(+)}, \quad
    h_\eta: \mathbb{R}^{m_u \times (T_f-1)} \rightarrow \mathbb{R}^{D \times (T_f-1)},
\end{equation}
mapping each input segment to a \( D \)-dimensional latent space, yielding the future latent input sequence \( Z_k^{(+)} \). For the emission history, the observation encoder maps the sequence \( P_k^{(-)} = [p_{k-T_p}, \dots, p_k] \), where \( T_p \) denotes the past horizon, according to
\begin{equation}
    g_\phi \big( P_k^{(-)} \big) = s_k, \quad
    g_\phi: \mathbb{R}^{m_p \times T_p} \rightarrow \mathbb{R}^D,
\end{equation}
producing the latent state \( s_k \) at index \( k \). The encoder parameterizations are represented by \( \eta \) and \( \phi \), respectively. Past and future emission sequences are denoted as \( P_k^{(-)} \) and \( P_k^{(+)} = [p_{k+1}, \dots, p_{k + T_f}] \), with future observations encoded analogously to obtain the latent states \( S_{k+1}^{(+)} \).

\textbf{Predictive Latent Dynamics:} From the latent domain, a predictor model evolves state representations auto-regressively
\begin{equation}
    f_\theta (s_k, z_k) = \Tilde{s}_{k+1}, \quad
    f_\theta: \mathbb{R}^{2D} \rightarrow \mathbb{R}^D,
\end{equation}
where \(\Tilde{s}_{k+1}\) approximates the next latent state. The mapping is iterated through the prediction horizon \( T_f \), forming the predicted latent sequence \( \Tilde{S}_{k+1}^{(+)} \).

\textbf{Loss Function Composition:} Robust latent modeling requires both compactness and structural informativeness in the representations. To this end, the latent space development is optimized using a composite loss incorporating VICReg (Variance-Invariance-Covariance-Regularization) and cross-covariance penalties~\cite{bardes_vicreg_2022}:

- The \textbf{Variance Loss} \( \mathcal{L}_v \) enforces a minimum spread of latent features
\begin{equation}
    \sigma(\mathbf{s}_{kd}) = \sqrt{\frac{1}{N-1} \sum_{n=1}^{N}
    \left( \mathbf{s}_{kd}[n] - \bar{\mathbf{s}}_{kd} \right)^2 + \epsilon_1 },
\end{equation}
with variance loss evaluated as
\begin{equation}
    \mathcal{L}_v(\mathbf{S}_k^{(+)}, \epsilon_1, \epsilon_2) = 
    \frac{1}{T_f D}
    \sum_{\kappa=1}^{T_f} \sum_{d=1}^{D}
    \frac{1}{\sigma(\mathbf{S}_k^{(+)}[:,\kappa,d], \epsilon_1) + \epsilon_2}
\end{equation}
where \( \epsilon_1 \) and \( \epsilon_2 \) are slack variables added to prevent numerical instabilities.

- The \textbf{Invariance Loss} \( \mathcal{L}_i \) penalizes deviations between simulated and actual latent state transitions:
\begin{align}
    \mathcal{L}_{i}(\mathbf{S}_{k+1}^{(+)}, \mathbf{\Tilde{S}}_{k+1}^{(+)}) =
    & \frac{1}{(T_f-1) N}
    \sum_{\kappa=1}^{T_f-1} \sum_{n=1}^{N} \nonumber \\
    & \left\| \mathbf{S}_{k+1}^{(+)}[\kappa,n] - \mathbf{\Tilde{S}}_{k+1}^{(+)}[\kappa,n] \right\|_2^2.
\end{align}

- The \textbf{Covariance Loss} \( \mathcal{L}_c \) mitigates feature redundancy by discouraging non-zero off-diagonal covariance terms:
\begin{equation}
    \mathcal{L}_c(\mathbf{S}_k^{(+)}) =
    \frac{1}{T_f (N-1)}
    \sum_{\kappa=1}^{T_f} \sum_{i=1}^{D} \sum_{j=i+1}^{D}
    \left(
    \mathbf{S}_k^{(+)}[\kappa] \mathbf{S}_k^{(+)\top}[\kappa]
    \right)_{i,j}.
\end{equation}

- The \textbf{Cross-covariance Loss} \( \mathcal{L}_{xc} \) incentivizes independence between latent input \( z_k \) and state \( s_k \) encodings
\begin{equation}
\label{eq:cross_covariance_loss}
    \mathcal{L}_{xc} = \frac{1}{D} \left| \frac{1}{N-1} \left( \mathbf{z}_k - \bar{\mathbf{z}}_k \right)^\top \left( \mathbf{s}_k - \bar{\mathbf{s}}_k \right) \right|_F^2
\end{equation}
using the squared Frobenius norm \( \|\cdot\|_F^2 \).

The principal hyperparameters and empirical performance metrics for the JEPA model are detailed in Table~\ref{tab:jepa_training}.

\begin{table}[h]
\centering
\begin{tabular}{p{0.6\linewidth}cp{0.3\linewidth}} 
\toprule
\textbf{Parameter} & \textbf{Value} \\
\midrule
    Encoder hidden dimensions           & [512, 256, 128] \\
    Predictor hidden dimensions         & [512, 512, 512] \\
    Input latent dimension              & 50 \\
    State latent dimension              & 50 \\
    Number of future timesteps ($T_f$)  & 2 \\
    Number of past timesteps ($T_p$)    & 10 \\
    Model size                          & \SI{8.03}{\mega\byte} \\
    Inference time (total)              & \SI{28.1537}{\second} \\ 
    Inference time (per timestep)       & \SI{14.0769}{\milli\second} \\
\bottomrule
\end{tabular}

\caption{Configuration and empirical characteristics of the JEPA model during training. 
The encoder and predictor network architectures are defined by their respective hidden dimensions. 
$T_p$ and $T_f$ denote the number of past and future timesteps used as model inputs and prediction horizons, respectively. The \textit{total inference time} refers to the total time required to perform inference of \num{2000} timesteps, and the \textit{per-timestep inference time} indicates the average time taken for each timestep during inference.}

\label{tab:jepa_training}
\end{table}

\begin{figure}[t]
  \begin{subfigure}
    \centering
    \includegraphics[width=0.43\textwidth]{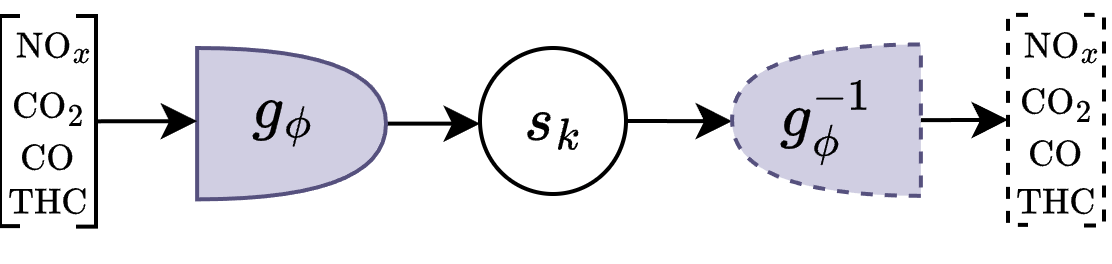}
    \label{fig:state_auto}
  \end{subfigure}
  \begin{subfigure}
    \centering
    \includegraphics[width=0.43\textwidth]{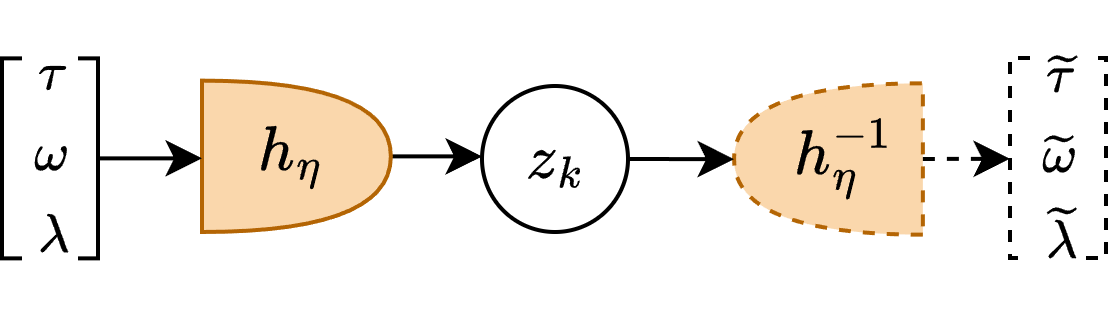}
    \label{fig:input_auto}
  \end{subfigure}
    \caption{Schematic representation of state and emission encoder–decoder. \\
    Diagram illustrating the relationships between engine operating variables ($\tau$, $\omega$, $\lambda$) and emission outputs (NO$_x$, CO$_2$, CO, THC), via their latent embeddings. Separate encoders ($g_\phi$ and $h_\eta$) transform input and emission sequences into their respective latent representations. The corresponding decoders ($g^{-1}_\phi$ and $h^{-1}_\eta$) reconstruct the original sequences from the latent space. This architecture enables reconstruction-based validation and interpretability of the learned latent dynamics, facilitating performance evaluation and closed-loop control design.
    }
    \label{fig:encoder_decoder}
\end{figure}

\textbf{Decoders for Domain Recovery:} To quantitatively evaluate latent space utility and facilitate interpretability, separate decoders were trained for emissions and input variables, as illustrated in Figure~\ref{fig:encoder_decoder}. Although external to the core JEPA module, these components reconstruct real-world states from latent representations using an MSE loss without affecting JEPA training.

\subsection{Model Compression}
\label{subsec:modelcompression}

Model compression encompasses a range of techniques that aim to reduce the size, computational cost, and energy consumption of deep neural networks while maintaining comparable predictive accuracy. It addresses the growing gap between increasingly large model architectures and the limited resources of deployment environments, such as embedded or edge devices. By reducing model complexity, compression enables more real-time inference and on-device learning while minimizing memory usage, latency, and power consumption.

\textbf{Pruning} is one of the most widely adopted compression techniques, removing redundant or less significant parameters from the network. Deep neural networks are often overparameterized, meaning that a substantial proportion of their weights contribute little to the final output. By systematically eliminating these parameters, pruning reduces the model’s storage footprint and accelerates inference. In practice, pruning strategies can either remove individual weights (\emph{unstructured pruning}) or entire channels, filters, or layers (\emph{structured pruning}). The latter maintains a regular tensor structure compatible with existing hardware accelerators, making it more practical for deployment.

\textbf{Quantization} is another prominent technique that reduces the numerical precision of model parameters and activations. Instead of high-precision 32-bit floating-point representations, the model’s weights are converted into lower-precision formats, such as 16-bit floating-point or 8-bit integer. This conversion substantially reduces memory footprint and computation time, as lower-precision arithmetic is significantly faster on modern processors. Post-training quantization methods directly convert a trained model to lower precision without additional retraining. In contrast, quantization-aware training exposes the model to quantization effects during training, enabling it to adapt and preserve accuracy more effectively. Quantization provides a substantial trade-off between performance efficiency and model fidelity, making it a core component of modern neural network deployment pipelines.

\vspace{-0.2cm}
\section{Results and Discussion}

\begin{figure*}[t!]
    \centering
    \includegraphics[width=0.95\textwidth]{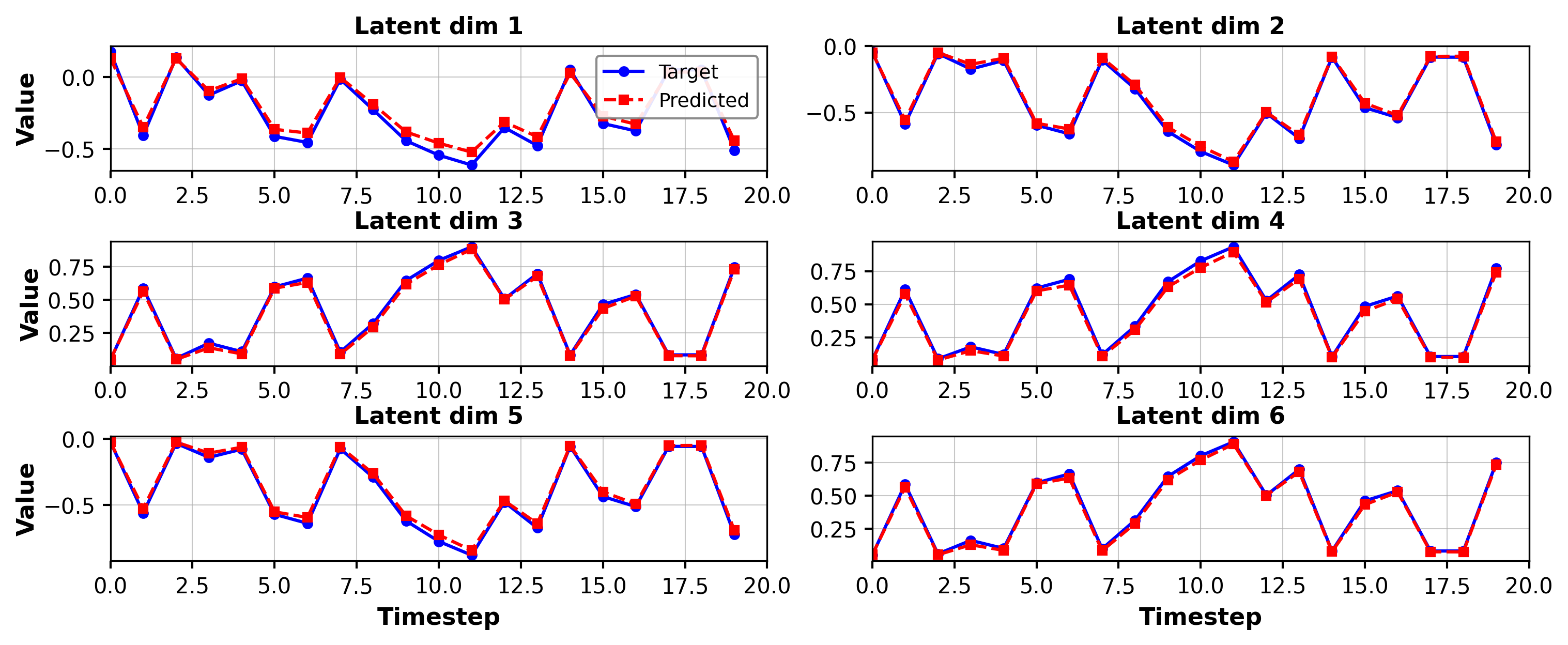}
    \caption{
        Validation of Latent Space Prediction Accuracy in the JEPA Model.\\
        The plots compare the predicted latent-state values (red) with the target latent-state values (blue) for the first six dimensions of the \num{50}-dimensional latent space over a \num{20}-step future horizon. The low error and close tracking between the predicted and target trajectories demonstrate that the predictor network successfully learned the temporal dynamics of the emission system within the latent representation.
    }
    \label{fig:latent_states_prediction}
\end{figure*}

\begin{figure*}[t!]
    \centering
    \includegraphics[width=\textwidth]{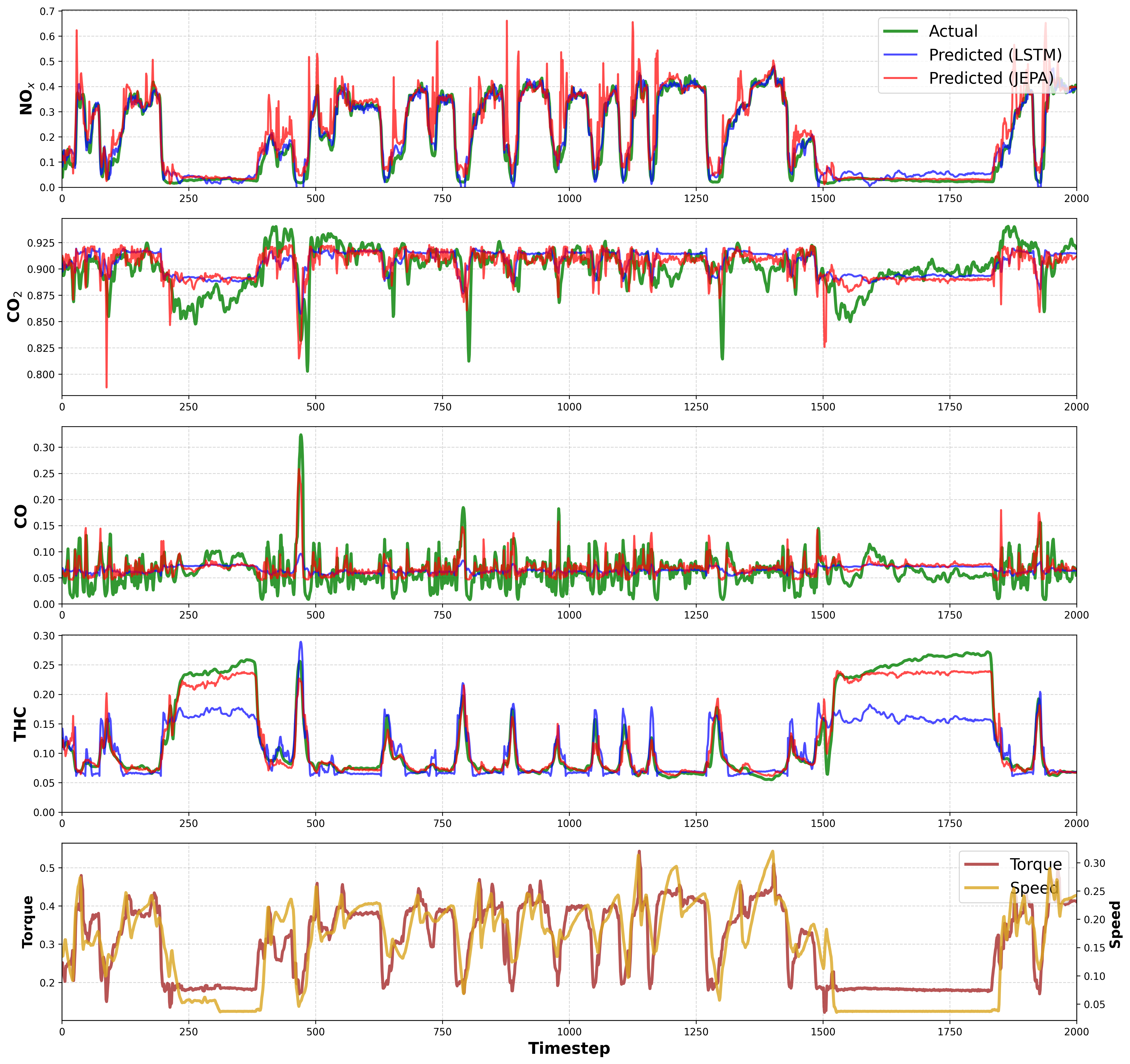}
    \caption{
        Comparative Analysis of Emission Prediction Performance: JEPA vs. LSTM.\\
        This figure shows a side-by-side comparison of the prediction accuracy for four key emission species: NO$_x$, CO$_2$, CO, and THC. The ground-truth measurements (Actual) are plotted in green, the LSTM model's predictions are in blue, and the JEPA model's predictions are in red. In the bottom diagram, the underlying test cycle is shown. All plot values are normalized to a range of 0 to 1 for consistent visualization.
    }
    \label{fig:emission_comp_jepa_lstm}
\end{figure*}

The performance of the JEPA framework is evaluated in two stages: first, by assessing its predictive accuracy within the latent space, and second, by comparing its real-world emission predictions against those of the LSTM baseline.

As detailed earlier, the JEPA training process aims to enable the predictor network ($f_\theta$) to accurately predict future latent states. The predictor receives the current latent state ($s_k$) from the state encoder ($g_\phi$) and the encoded input ($z_k$), and recursively predicts a sequence of future latent states. The primary training objective is to minimize the discrepancy between the predicted and target latent states, which are obtained by encoding the actual future emission measurements. A low latent-space prediction error indicates that the predictor has successfully internalized the underlying dynamics of the emission system. Figure~\ref{fig:latent_states_prediction} provides a representative validation of this process, illustrating a close correspondence between the predicted and target values for the first six dimensions of the 50-dimensional latent space. This alignment confirms that the predictor accurately models the system's temporal evolution.

Although latent space performance indicates the model's learning capability, it can only be validated in the original data domain. For this purpose, the separately trained decoders transform JEPA's latent predictions into physical emission values. This allows for a direct, one-to-one comparison with the LSTM model. Both models were tasked with predicting emissions over a \num{2000} timestep test cycle, and the comparative results are presented in Figure~\ref{fig:emission_comp_jepa_lstm}.

While the LSTM model captures general trends in emission dynamics, the JEPA model provides a superior fit to the data. This is particularly evident during rapid transient changes in CO$_2$ and CO concentrations, mainly when steep torque ramps occur at moderate to high engine speed. In these regions, the LSTM model tends to produce an averaged, smoothed response with a small phase lag, whereas the JEPA model more closely follows both the fast excursions and the slower background dynamics. In quasi-steady operation (speed nearly constant while torque varies slowly), both models remain aligned for CO$_2$. In contrast, during brief tip-in/tip-out events or torque reversals, JEPA tracks the short dips and recoveries with less delay.

For NO$_x$ emissions, the JEPA model more accurately captures the magnitude and timing of sharp peaks that arise during fast load increases. In contrast, the LSTM consistently underestimates these peak values and smooths their onset. A trade-off is observed on very short impulses. JEPA occasionally overshoots and introduces isolated spikes, while the LSTM produces smoother trajectories with fewer spurious transients during extended low-load or coast segments (low torque, nearly constant speed). Furthermore, the LSTM model exhibits a noticeable phase shift and damped pulse amplitudes under repeated positive load transients at roughly constant speed in the THC predictions. The JEPA model, in contrast, shows minimal shift and preserves the crest levels more faithfully. 

These observations indicate that JEPA’s predictive latent dynamics are more sensitive to torque rate, thereby improving fidelity in emission-relevant transients across the speed–torque map. At the same time, the LSTM remains competitive in quasi-steady operation but tends to underestimate peaks and exhibit lag during rapid load changes. Overall, JEPA exhibits a higher bandwidth response, enabling it to follow fast transient excursions more faithfully across multiple species. In a small number of isolated, very short-duration events, this sensitivity can manifest as slight overshoot, suggesting that additional temporal regularization or robust error formulations could further improve spike suppression without sacrificing transient fidelity.

\begin{figure*}[t!]
    \centering
    \includegraphics[width=\textwidth]{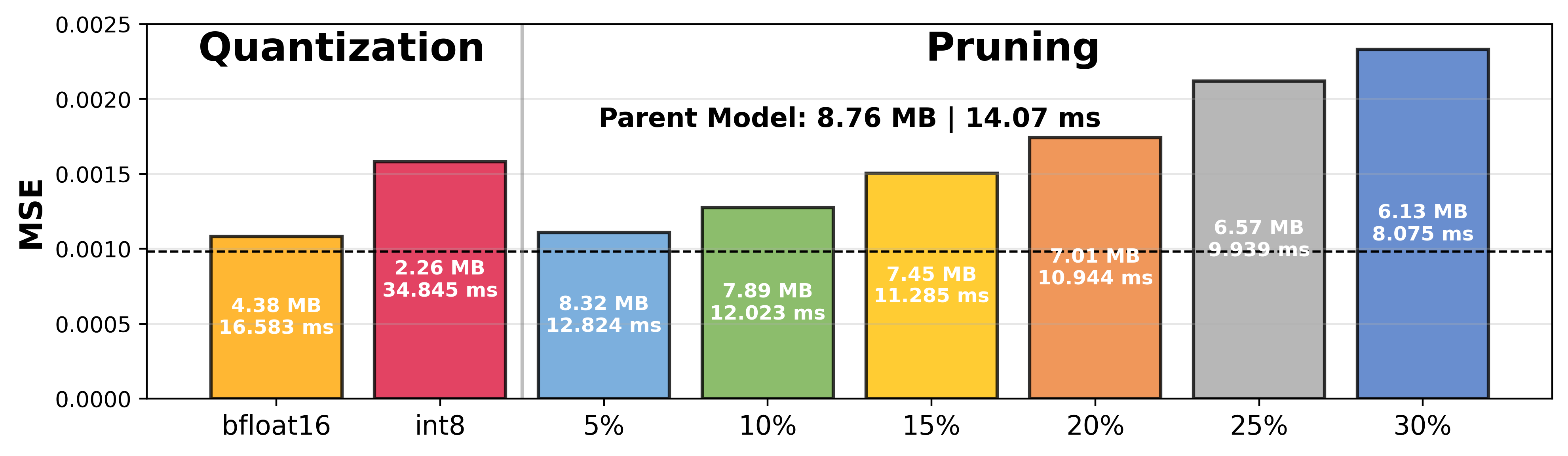}
    \caption{
        Impact of Quantization and Pruning on JEPA Model Performance.\\
        The figure illustrates the trade-offs between model accuracy (MSE), model size (MB), and inference time (ms) when applying different compression techniques to the baseline parent model. The left panel shows the effects of quantizing the model to bfloat16 and INT8 precisions. The right panel shows the impact of structured pruning at ratios ranging from 5\% to 30\%. Each bar shows the resulting MSE, with annotations indicating the model size and inference latency. The dashed line represents the MSE of the original, uncompressed parent model, providing a clear reference for performance degradation.
    }
    \label{fig:compressed_model}
\end{figure*}

\textbf{Impact of Model Compression on JEPA}: To evaluate the feasibility of deploying the JEPA model in resource-constrained environments, its performance was benchmarked under two distinct compression strategies: structured pruning and post-training quantization. The parent (uncompressed) model serves as the baseline, with a size of \SI{8.76}{\mega\byte} and an inference time of \SI{14.07}{\milli\second}. First, structured pruning was applied at 5\%, 10\%, 15\%, 20\%, and 30\%, yielding multiple models with 5\% pruning intervals. Second, the parent model was quantized from its native float32 precision to bfloat16 and INT8 formats. Figure~\ref{fig:compressed_model} illustrates the impact of these techniques on model size, inference time, and predictive accuracy, as measured by Mean Squared Error (MSE).

The results reveal a clear trade-off between compression level and model performance. As illustrated in the pruning results, increasing the pruning ratio progressively reduces the model size and inference time. For instance, pruning 30\% of the model decreases its size to \SI{6.13}{\mega\byte} and speeds up inference to \SI{8.075}{\milli\second}. However, this efficiency gain is accompanied by a significant loss of accuracy, with MSE more than doubling from approximately \num{0.001} to over \num{0.0022}. This noticeable increase in error beyond a 20\% pruning threshold suggests that higher pruning ratios begin to remove structurally critical parameters, which are essential for the model's predictive capabilities. The quantization results present a different set of trade-offs. Converting the model to bfloat16 halves the model size to \SI{4.38}{\mega\byte} and yields a modest increase in MSE, indicating a viable strategy for moderate compression with minimal accuracy loss. In contrast, quantizing to INT8 achieves the most significant size reduction to \SI{2.26}{\mega\byte} but at the cost of a drastic increase in both MSE and, unexpectedly, inference time (\SI{34.845}{\milli\second}). This counterintuitive increase in latency is likely due to hardware inefficiencies: operations for INT8 or bfloat16 arithmetic are less well optimized than those for floating-point formats on the target platform, resulting in computational overhead that offsets the benefits of a smaller model size.

\vspace{-0.2cm}

\section{Conclusion and Future Work}

This work introduced a latent-space modeling framework based on JEPA for transient emission prediction and systematically compared its performance against an LSTM baseline. Using real-world driving trajectories replayed on a dynamometer test bench, the study demonstrated that JEPA learns a compact, structured state representation that captures the relationships between engine operating variables and emission responses. In demanding transient regimes characterized by rapid torque and speed changes, JEPA produced predictions with closer peak tracking and a smaller phase shift than LSTM, highlighting its superior capacity to model both fast and slow nonlinear dynamics. The latent predictor effectively reproduced temporal emission dynamics with low reconstruction error, and decoders successfully recovered physically meaningful signals, confirming that the embedding retained control-relevant structure. Model compression analysis revealed that structured pruning and quantization can significantly reduce JEPA's computational footprint. Moderate bfloat16 quantization halved model size with only marginal accuracy degradation, while aggressive techniques, such as deep structured pruning or INT8 quantization, introduced pronounced efficiency-performance trade-offs, underscoring the need for carefully calibrated, deployment-specific compression strategies. This study validates JEPA's potential as a high-fidelity tool for modeling transient emissions. Its ability to learn robust, low-dimensional representations marks a meaningful advance over conventional recurrent architectures and opens new avenues for embedding-based control.

The study found two fundamental limitations in the experimental setup. First, the input feature set was restricted to three ECU signals: engine torque, speed, and air–fuel ratio. While JEPA effectively exploited these signals, this limited observability may constrain accuracy during sharp transients, during which unobserved engine conditions dominate emission dynamics. Future work should explore additional sensor inputs and derived variables, including manifold pressures, air mass flow, exhaust temperature, fuel rate, ignition timing, boost pressure, and vehicle signals (such as gear and speed), to increase observability, enrich latent embeddings, and improve the fidelity of transient predictions. Second, JEPA's multi-component architecture, comprising separate encoders, a latent predictor, and decoders, incurs higher inference latency and memory demand than the monolithic LSTM. Addressing this requires hardware-aware compression strategies that respect automotive-grade real-time requirements. Future research will prioritize component-specific pruning and quantization schemes that preserve encoder fidelity while aggressively compressing downstream components, balancing efficiency with task-critical accuracy.

Beyond compression, combining JEPA with physics-based modeling to create hybrid gray-box architectures represents an important research direction. Physics-informed regularizers that enforce mass-balance consistency, positivity constraints, or monotonicity in latent dynamics can guide learning toward compact, predictive, and physically interpretable representations, making them robust to extrapolation. Constraining latent dynamics to respect fundamental physical laws improves identifiability, out-of-distribution generalization, and interpretability without sacrificing computational efficiency. Richer sensor integration, hardware-aware compression, and physics-guided latent learning provide a coherent roadmap toward deployable, high-fidelity emission-prediction systems for next-generation powertrain control.

\vspace{-0.2cm}

\section{Contact Information}
\vspace{-0.2cm}

Tobias Gehra, PhD \newline
RPTU Kaiserslautern-Landau, Germany \newline
Institute of Vehicle Propulsion Systems (LAF) \newline
tobias.gehra@mv.rptu.de

Ganesh Sundaram, M.Sc. \newline
RPTU Kaiserslautern-Landau, Germany\newline
Institute of Electromobility (LEM) \newline
ganesh.sundaram@rptu.de

\section{Acknowledgments}
Part of the research presented in this paper was conducted within the project \textit{ML-MoRE} (Maschinelles Lernen für die Modellierung und Regelung der Emissionen von Hybridfahrzeugen in Realfahrzyklen). This project was funded by the German Federal Ministry of Education and Research (BMBF) under grant number 01-S20007A. The authors are solely responsible for the content of this publication and gratefully acknowledge the financial support provided.

\begin{figure}[H]
    \flushleft
    \includegraphics[width=0.6\columnwidth]{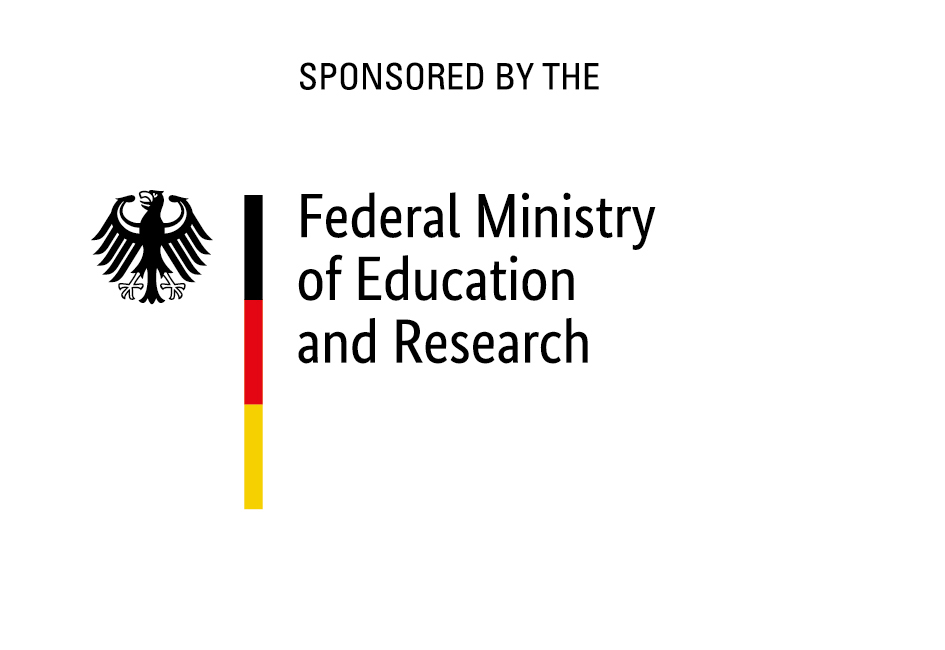}
    \label{fig:Logo}
\end{figure}

Furthermore, the Deutsche Forschungsgemeinschaft (DFG, German Research Foundation) funded part of the measurement equipment used in this study under project number 463549337. The authors express their gratitude for this support.

The authors would also like to thank the project partners, KST-Motorenversuch and RA Consulting, for their collaboration and technical support throughout the ML-MoRE project.

\bibliographystyle{ieeetr} 
\bibliography{main}

\end{document}